\documentclass[letterpaper,letter,twocolumn]{jpsj3}
\usepackage{txfonts,bm,color}

\title{Optical Selection Rules in Spin-Orbit Coupled Systems on Honeycomb Lattice}

\author{Yuki Yanagi\thanks{y\_yanagi@meiji.ac.jp} and Hiroaki Kusunose}
\inst{Department of Physics, Meiji University, Kawasaki 214-8571, Japan}

\abst{
We investigate the optical properties in a minimal model of spin-orbit coupled systems on honeycomb lattice at half-filling.
The absorption of the circularly polarized light in the charge and antiferromagnetic ordered states is particularly studied, and the spin-valley selective excitation depending on the handedness of the light is found.
It is shown that the optical selection rules plainly differ with the type of broken symmetries and topological properties of the electronic states.
Consequently, the abrupt change of spin-valley selectivity in the optical absorption occurs at the topological transition caused by changing the magnitude of the order parameters.
}

\begin{document}
\maketitle


Spin-orbit coupling in noncentrosymmetric lattice structures has evoked vivid interest since it plays an important role in realizing topological insulators, unconventional superconductivity, multiferroics, and so on~\cite{Hasan10,Xiao_1,Yao,Xiao_2,Ezawa_1,Saito15,Wakatsuki17,Khomskii09,Planes14,Nagaosa12,Sinitsyn08}.
Instead of considering such built-in noncentrosymmetric lattice structures, there has been a growing interest in the study of novel quantum phases in the centrosymmetric lattices with local asymmetry at the correlated sites, e.g. the zigzag, honeycomb, and bilayer lattice structures~\cite{Katsura05,Mostovoy06,Arima11,Khanh16,Fang14,Fu15,Ederer,Spaldin_1,Spaldin_2,Li,Maruyama,Sigrist,Yanase,Hayami_1,Hayami_2,Hayami_5,Hitomi,Yoshida,Hayami_3,Sumita,Nii17,Hayami_4,Nakatsuji15,Suzuki17}.
In such systems, the antisymmetric spin-orbit coupling (ASOC), which is hidden in a staggered form, can be a source of intriguing physical phenomena as similar to the built-in noncentrosymmetric systems.

Previous studies have shown that the staggered spin-, charge- and orbital-orderings break spontaneously the spatial inversion symmetry accompanying the ferroic component of the odd-parity multipole moment~\cite{Ederer,Spaldin_1,Spaldin_2}, and the uniform ASOC hidden in the normal state is activated~\cite{Hayami_5}.
In Refs.~\citen{Hayami_2} and \citen{Hayami_5}, Hayami \textit{et al}. have introduced a minimal two-orbital model on the honeycomb lattice including the spin-, charge-, orbital- and valley-degrees of freedom to elucidate the essential physics in systems with the local asymmetry.
The effects of spontaneous parity breaking on the electronic structures and static response functions in the two-orbital model have been intensively studied.
It has been revealed that the rich magnetoelectric effects arise from the interplay of multiple electronic degrees of freedom~\cite{Hayami_2,Hayami_5}.

In this Letter, we study the optical properties of the two-orbital model introduced in Refs.~\citen{Hayami_2} and \citen{Hayami_5}  at half-filling by calculating the optical conductivity.
It will be demonstrated that the spin-valley dependent circular dichroism appears in the charge and antiferromagnetic ordered states and that the optical selection rules are different in each states.
We will also find that there is close relation between optical selection rules and topological properties.
This leads to abrupt change of the selection rules accompanied by the topological transition.

\begin{figure}[t]
\begin{center}
\includegraphics[width=8cm]{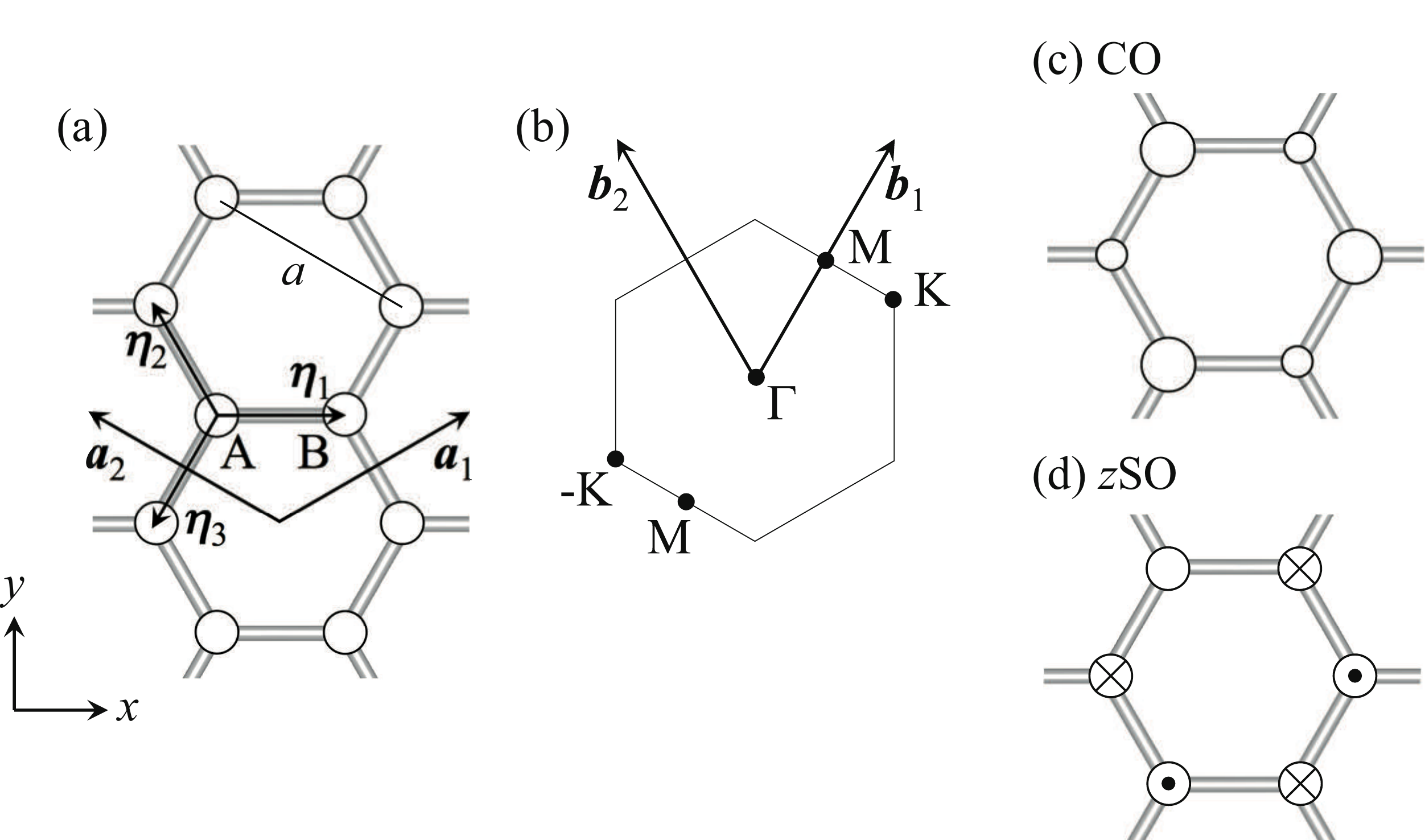}
\caption{
(a) Schematic illustration of the honeycomb lattice with the definitions of the primitive translation vectors $\bm{a}_i$ ($i=1,2$) and the nearest-neighbor bonds from A to B sites $\bm{\eta}_i$ ($i=1,2,3$).
(b) First Brillouin zone of the honeycomb lattice with the reciprocal lattice vectors $\bm{b}_i$ ($i=1,2$), and the high-symmetry points.
(c) The charge-ordered state (CO) and (d) the spin-ordered state along $z$ axis ($z$SO) to be considered in this work.  
}
\label{fig_lattice}
\end{center}
\end{figure}

The two-orbital model  is given by\cite{Hayami_2,Hayami_5},
\begin{align}
\mathcal{H}=\mathcal{H}_0+\mathcal{H}_{\mathrm{SOC}}+\mathcal{H}_{\mathrm{MF}},
\label{eq_H_tot}
\end{align}
where the first, second and third terms represent the kinetic, spin-orbit coupling and molecular-field terms, respectively.
The lattice structure and corresponding Brillouin zone together with lattice vectors are shown in Figs.~\ref{fig_lattice}(a) and (b).
The kinetic term $\mathcal{H}_0$ is expressed as
\begin{align}
\mathcal{H}_0 = \sum_{\bm{k} ss' mm' \sigma} \left[\hat{H}_0(\bm{k})\right]_{sm,s'm'}c^{\dagger}_{\bm{k}s m\sigma}c^{}_{\bm{k}s' m'\sigma},
\end{align}
where $c^{(\dagger)}_{\bm{k}s m\sigma}$ is the annihilation (creation) operator for the $d$-electron with wave vector $\bm{k}$, orbital magnetic quantum number $m=\pm 1$, spin $\sigma=\uparrow$, $\downarrow$ on the sublattice $s=\mathrm{A}$, $\mathrm{B}$.
The matrix elements of $\hat{H}_0(\bm{k})$ are given as follows,
\begin{align}
&
\left[\hat{H}_0(\bm{k})\right]_{\mathrm{A}m,\mathrm{B}m}= -t_0 \gamma_{0\bm{k}},
\quad
\left[\hat{H}_0(\bm{k})\right]_{\mathrm{A}m,\mathrm{B}-m}= -t_1 \gamma_{m\bm{k}},
\cr
&
\left[\hat{H}_0(\bm{k})\right]_{\mathrm{B} m,\mathrm{A} m' }=\left[\hat{H}_0(\bm{k})\right]^*_{\mathrm{A} m',\mathrm{B} m }.
\end{align}
Here, $t_0$ and $t_1$ are the nearest-neighbor intra- and inter-orbital hoppings, respectively, and $\bm{k}$-dependence of the $\gamma_{n\bm{k}}$, $(n=0,\pm1)$ is given by
$\gamma_{n\bm{k}}=\sum_{i=1}^3 \omega^{(i-1)n}e^{i\bm{k}\cdot \bm{\eta}_i} =\gamma^*_{-n-\bm{k}}$, 
where $\omega=e^{i2\pi/3}$, and $\bm{\eta}_{i}$ are the nearest-neighbor bonds.

The spin-orbit coupling term $\mathcal{H}_\mathrm{SOC}$ is given by
\begin{align}
\mathcal{H}_\mathrm{SOC} = \frac{\lambda}{2} \sum_{\bm{k}sm\sigma} m\sigma c^{\dagger}_{\bm{k}sm\sigma}c^{}_{\bm{k}s m\sigma},  
\end{align}
where  $\lambda$ represents the strength of the spin-orbit coupling.
In our model, the spin-orbit coupling is Ising-type because  only  $m=\pm 1$ orbitals are included.

\begin{figure}[t]
\begin{center}
\includegraphics[width=8cm]{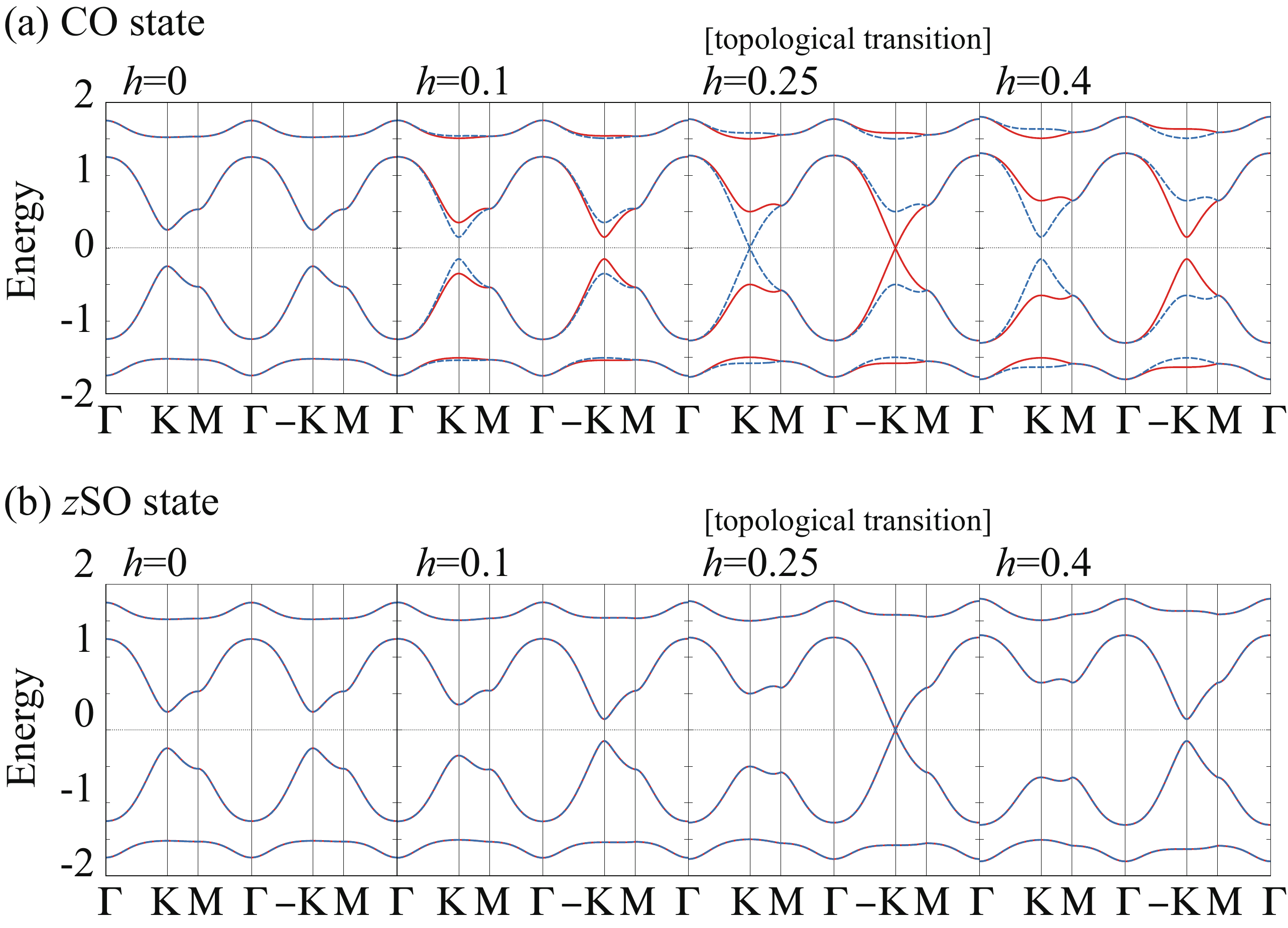} 
\caption{
(Color online) Band structures in (a) the CO state and (b) the $z$SO state.
Solid (red) and dashed (blue) lines represent the bands for spin-$\uparrow$ and -$\downarrow$ electrons.
}
\label{fig_band}
\end{center}
\end{figure}

The third term in Eq.~(\ref{eq_H_tot}) describes the molecular field due to the spontaneous symmetry breaking, and is given by
\begin{align}
\mathcal{H}_\mathrm{MF} = -h \sum_{\bm{k}s \sigma\sigma'mm'} p(s)\left[\sigma_\alpha \right]_{\sigma\sigma'}\left[\tau_\beta \right]_{mm'}c^{\dagger}_{\bm{k}s m\sigma}c^{}_{\bm{k}s m'\sigma'},
\label{ham-mf}
\end{align}
where $\sigma_{\alpha}$ and $\tau_\beta$ ($\alpha,\beta=0,x,y,z$) are the unit and Pauli matrices in the spin and orbital spaces, respectively.
In the present study, we focus on the charge ordered state ($\alpha=\beta=0$: CO) and spin ordered state along $z$ axis ($\alpha=z$, $\beta=0$: $z$SO) of the staggered type, which is ensured by the factor $p(s)=+1$ $(-1)$ for $s=\mathrm{A}$ $(\mathrm{B})$.
Note that the spin $\sigma$ is a good quantum number in the CO and $z$SO states, which are depicted  in Fig.~\ref{fig_lattice}.

The single-particle eigenenergy $\varepsilon_{\bm{k}\sigma\nu}$ and corresponding eigenvector $|\bm{k}\sigma\nu\rangle$ are obtained by diagonalizing the Hamiltonian in Eq.~(\ref{eq_H_tot}) at each $\bm{k}$, where we label the bands $\nu~(=1,2,3,4)$ in ascending order of energy.
It should be noted that by the following particle-hole (PH) transformation~\cite{Shiba},
\begin{align}
c^{\dagger}_{\bm{k}sm\uparrow}\rightarrow c^{\dagger}_{\bm{k}sm\uparrow},
\quad
c^{\dagger}_{\bm{k}sm\downarrow}\rightarrow p(s)c^{}_{\bm{-k}s-m\downarrow},
\label{shibatrans}
\end{align}
the CO state is converted to the $z$SO state, and simultaneously $\varepsilon_{\bm{k}\uparrow\nu}\to\varepsilon_{\bm{k}\uparrow\nu}$ and $\varepsilon_{\bm{k}\downarrow\nu}\to-\varepsilon_{-\bm{k}\downarrow 5-\nu}$.
For simplicity, we set $t_0=t_1=\lambda=0.5$ hereafter, since the essential conclusions are not altered with other choices.

Figure~\ref{fig_band} shows the energy bands $\varepsilon_{\bm{k}\sigma\nu}$ in the normal, CO and $z$SO states.
In the normal state, time-reversal symmetry ($\theta$), inversion symmetry ($I$), and combined symmetry of $\theta$ and $I$ ($I\theta$) are maintained.
Then, the relation $\varepsilon_{\bm{k}\sigma\nu}=\varepsilon_{-\bm{k} \sigma\nu}=\varepsilon_{\bm{k}-\sigma\nu}$ holds as shown in Fig.~\ref{fig_band}.
The top (bottom) of the valence (conduction) band is located on the $\pm$K-points [$\bm{k}=\pm\bm{K}=\pm(2\bm{b}_1-\bm{b}_2)/3$] and  the system has a valley degree of freedom.
The excitations at the valleys play significant roles to determine the optical properties at low energy.
The spontaneous breaking of spatial parity induces inequivalence of K- and $-$K-points and makes the valley-dependent excitation possible~\cite{Xiao_1,Yao,Xiao_2,Ezawa_1,Li} as will be shown later.

When the molecular field is turned on, the symmetry is lowered and the band structure is deformed.
In the case of the CO state, $\theta$ is preserved while $I$ and $I\theta$ are broken.
This leads to the spin splitting of the band structure~\cite{Hayami_5}, \textit{i.e.}, $\varepsilon_{\bm{k}\sigma\nu}=  \varepsilon_{-\bm{k} -\sigma\nu}$ while $\varepsilon_{\bm{k}\sigma\nu}\neq\varepsilon_{\bm{k}-\sigma\nu}$.
The smallest direct gap is given by $\Delta_1=|2h-\lambda|$, and gets smaller with increasing the CO molecular field $h$ for $h<h_{\rm c}=\lambda/2$ ($=0.25$) and closes at $h=h_{\rm c}$ [see Fig.~\ref{fig_band}(a)].
With further increase of $h$, the gap reopens and gets larger for $h > h_{\rm c}$.

On the other hand, under the $z$SO molecular field,  $I\theta$ is preserved but $I$ and $\theta$ are broken.
Reflecting the different symmetry breaking from the CO state, the distinct band structures are obtained as shown in Fig.~\ref{fig_band}(b), where the relation $\varepsilon_{\bm{k}\sigma\nu}=\varepsilon_{\bm{k}-\sigma\nu}\neq\varepsilon_{-\bm{k}\sigma\nu}$ holds in the $z$SO state.
It exhibits the valley splitting~\cite{Hayami_5}, where the magnitude of the direct gap at $-$K-point, $\Delta_1=|2h-\lambda|$, is  smaller than that at K-point, $\Delta_2=2h+\lambda$.
Note that the gap closing point $h_{\rm c}$ for the $z$SO state is exactly the same as that for the CO state owing to the PH transformation of Eq.~(\ref{shibatrans}).

Now, we discuss topological properties in the CO and $z$SO states since those are closely related to the optical selection rules as will be discussed later.
As shown in the previous study~\cite{Hayami_5}, in the CO state, spin Hall conductivity $\sigma_\mathrm{sh}$ is quantized as $\sigma_\mathrm{sh}=2$ up to $h=h_{\rm c}$ while $\sigma_\mathrm{sh}=0$ for $h>h_{\rm c}$.
It is confirmed that the $h$-dependence of $\sigma_\mathrm{sh}$ in the $z$SO state is same as that in the CO state, also due to the PH transformation.
That is to say, quantization of $\sigma_\mathrm{sh}$ is realized for $h<h_{\rm c}$ also in the $z$SO state even though the time-reversal symmetry is broken.
This indicates that the topological phase for $h<h_{\rm c}$ in the CO and $z$SO state is so-called spin Chern insulator~\cite{Ezawa_2} rather than the quantum spin Hall insulator~\cite{Kane}.
It is essential that the different spin sectors, \textit{i.e.,} $\sigma=\uparrow$ and $\downarrow$ are decoupled in the phases we concern in the present study.
One of the characteristics of spin Chern insulator is the possible topological transition without gap closing in addition to the conventional one with the band gap closing.
In fact, it is verified that by switching on the transverse magnetic field, or in the case of $\alpha=x$, $\beta=0$ in Eq.~(\ref{ham-mf}) to mixture the different spin sectors, the spin Hall conductivity is not quantized and decreases continuously from $\sigma_\mathrm{sh}=2$ with increasing $h$ without gap closing.

Let us turn our attention to  optical response.
In order to clarify the optical properties of the present model, we calculate the optical conductivity $\sigma_{+(-)}(\omega)$ with use of the Kubo formula, which is proportional to the probability of right (left) circularly polarized light absorption with frequency $\omega$.
According to the linear response theory, the spin-resolved optical conductivity $\sigma^{\sigma\sigma'}_{\pm}(\omega)$ [$\sigma_{\pm}(\omega)=\sum_{\sigma\sigma'}\sigma_{\pm}^{\sigma\sigma'}(\omega)$] is given by 
\begin{align}
&
\sigma^{\sigma\sigma'}_{\pm}(\omega)= \mathrm{Re}\,
\frac{Q^{\sigma\sigma'}_{\pm}(\omega)-Q^{\sigma\sigma'}_\pm(0)}{i\omega V},
\quad
Q^{\sigma\sigma'}_{\pm}(\omega)=
\langle\langle j^{\sigma}_{\mp}; j^{\sigma'}_{\pm} \rangle\rangle_\omega,
\end{align}
where $V=N\sqrt{3}a^2/2$ is the volume.
The operator $j^{\sigma}_{\pm}$ is defined as $j^{\sigma}_{x}\pm i j^{\sigma}_{y}$, where $j^{\sigma}_{r}$ $(r=x,y)$ is a current operator along $r$-direction for spin $\sigma$ given by 
\begin{align}
j^{\sigma}_{r} = \sum_{\bm{k}ss'mm'} \left[\frac{\partial \hat{H}_0(\bm{k})}{\partial k_r} \right]_{sm,s'm'}c^{\dagger}_{\bm{k}sm\sigma}c^{}_{\bm{k}s' m'\sigma}.
\end{align}
Since the spin $\sigma$ is a good quantum number in the CO and $z$SO states, $\sigma_\pm^{\sigma\sigma'}(\omega)$ becomes spin diagonal, \textit{i.e.,} $\sigma_\pm^{\sigma\sigma'}(\omega)=\sigma_\pm^{\sigma}(\omega)\delta_{\sigma\sigma'}$.
Then, $\sigma_{+(-)}^{\sigma}(\omega)$ represents the probability of the excitation for spin-$\sigma$ electron with respect to the absorption of the right (left) circularly polarized light.
The spin-resolved optical conductivity $\sigma_\pm^{\sigma}(\omega)$ is explicitly written as
\begin{align}
\sigma^{\sigma}_{\pm}(\omega)
=\mathrm{Re}\,\frac{1}{iV}\sum_{\bm{k}\nu\nu'}\frac{| \langle  \bm{k}\sigma\nu'| j_{\pm}^{\sigma} |\bm{k} \sigma \nu \rangle|^2}{\varepsilon_{\bm{k}\sigma\nu}- \varepsilon_{\bm{k}\sigma\nu'}}
\frac{ f(\varepsilon_{\bm{k}\sigma\nu})-f(\varepsilon_{\bm{k}\sigma\nu'}) }  {\omega+\varepsilon_{\bm{k}\sigma\nu}- \varepsilon_{\bm{k}\sigma\nu'} +i\gamma},
\end{align}
where $f(\varepsilon)=1/(e^{\varepsilon/T}+1)$ is the Fermi distribution function, $T$ is the temperature and $\gamma$ is a  small positive constant.
In our numerical calculation, we set $T=\gamma=0.01$.
 
\begin{figure}[t]
\begin{center}
\includegraphics[width=8cm]{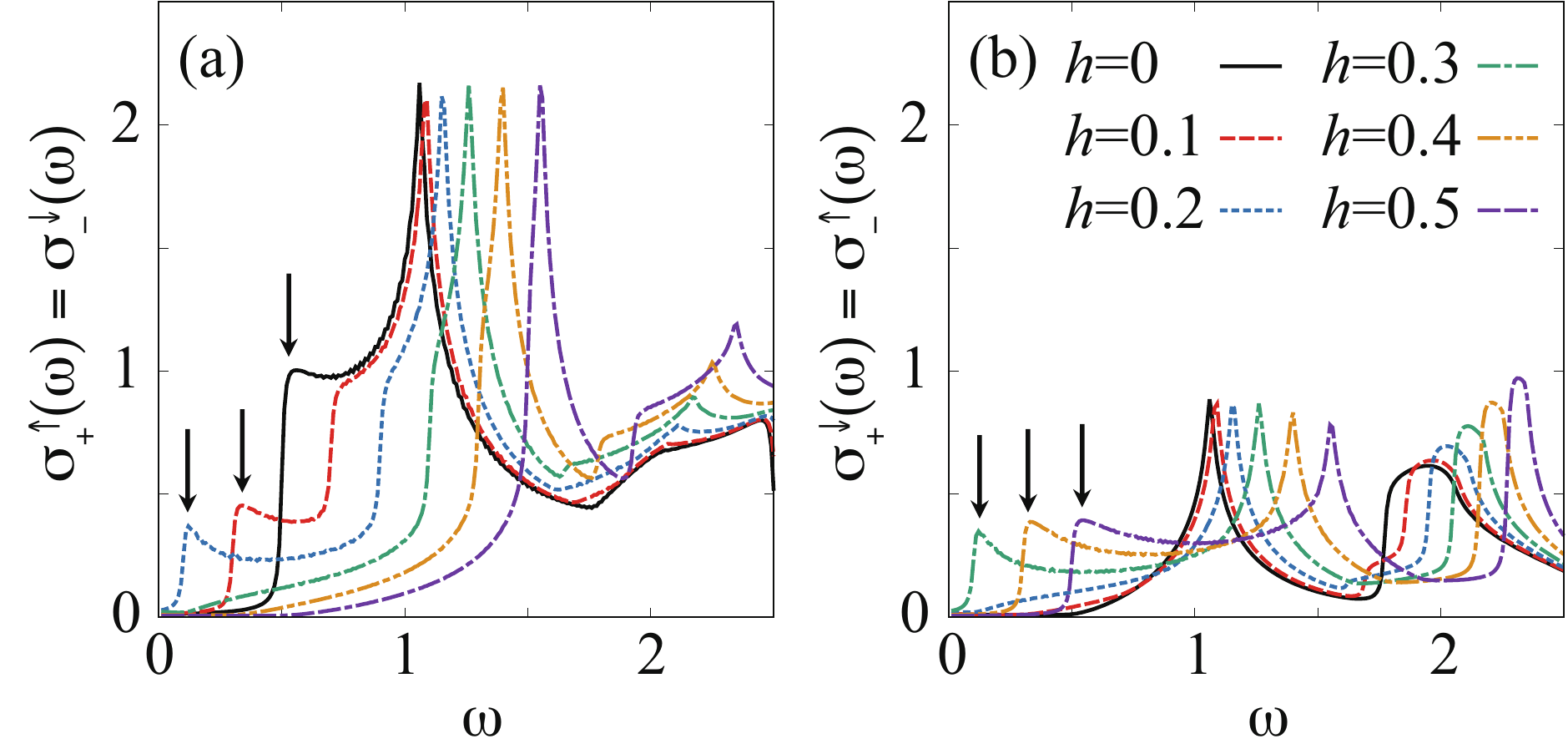}
\caption{
(Color online) Optical conductivity for $h\le 0.5$ as a function of $\omega$.
(a) $\sigma_+^{\uparrow}(\omega)=\sigma_-^{\downarrow}(\omega)$ and (b) $\sigma_+^{\downarrow}(\omega)=\sigma_-^{\uparrow}(\omega)$.
The peak position at $\omega=\Delta_1$ is denoted by the arrows.
}
\label{fig_sigma}
\end{center}
\end{figure}

Before going to the numerical results, we discuss several symmetry properties of $\sigma^{\sigma}_\pm(\omega)$.
In the CO ($z$SO) state, the Hamiltonian has $\theta$- ($I\theta$-) symmetry as explained before, while the current operator is transformed as $\theta j^{\sigma}_{\pm} \theta^{-1}= -j^{-\sigma}_{\mp}$ ($ I\theta j^{\sigma}_{\pm} (I\theta)^{-1} = j^{-\sigma}_{\mp}$).
This leads to the relation of the optical conductivity as $\sigma^{\uparrow}_{\pm}(\omega)=\sigma^{\downarrow}_{\mp}(\omega)$ in the normal and both ordered states.
Moreover, by the PH transformation, Eq.~(\ref{shibatrans}), the CO state is converted into the $z$SO state and simultaneously the current operator is transformed as $j^{\downarrow}_{\pm} \rightarrow -j^{\downarrow}_{\pm}$.
Therefore, the optical conductivity $\sigma^{\sigma}_{\pm}(\omega)$ in the CO state exactly coincides with that in the $z$SO state.

\begin{figure}[t]
\begin{center}
\includegraphics[width=8cm]{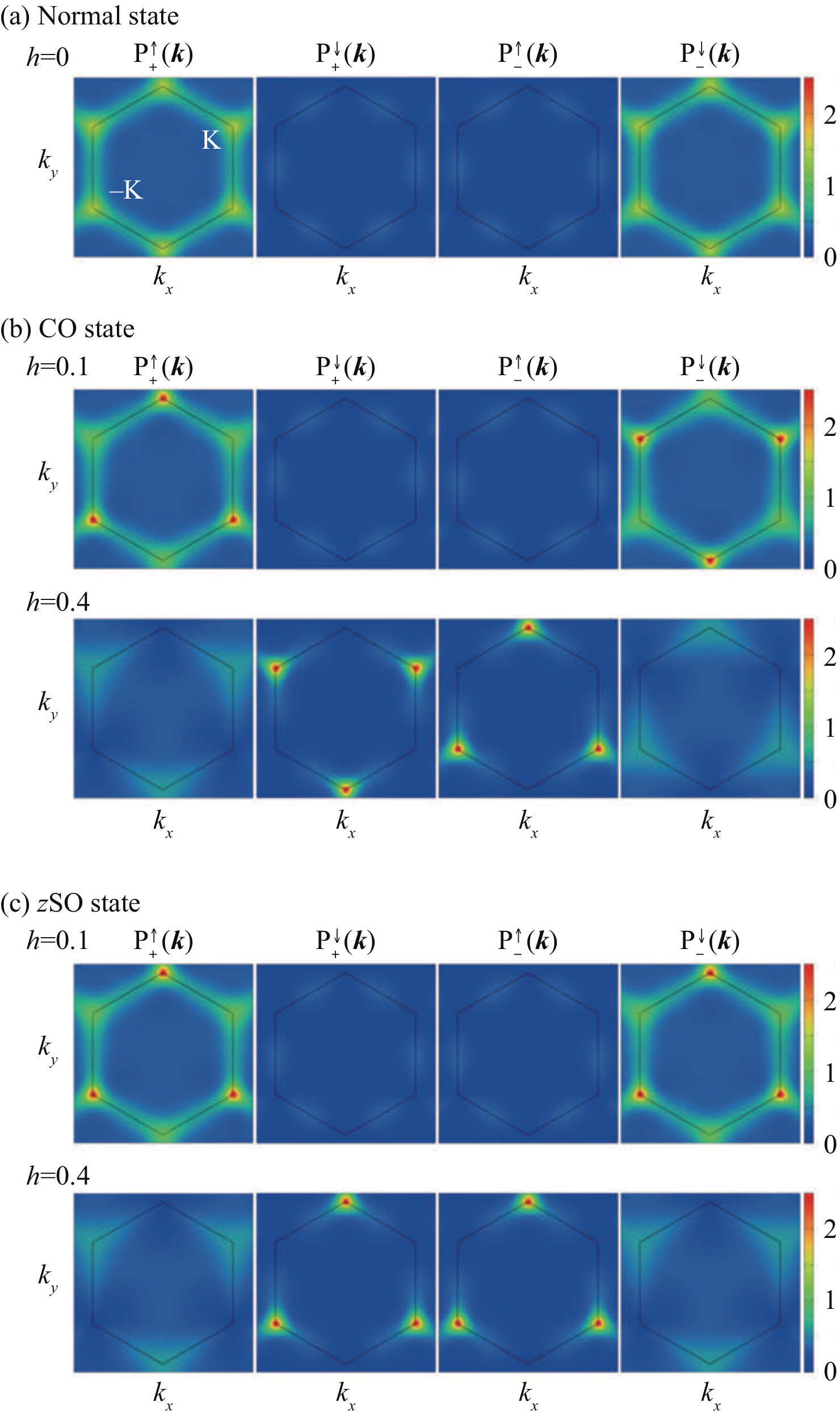} 
\caption{
(Color online) Momentum-resolved intensities of the optical absorption $P^{\sigma}_{\pm}(\bm{k})$ (a) in the normal, (b) CO, and (c) $z$SO states.
First Brillouin zone is indicated by the solid line.
}
\label{fig_intensity}
\end{center}
\end{figure}

Figures~\ref{fig_sigma}(a) and (b) show the spin-resolved optical conductivity $\sigma^{\sigma}_{\pm}(\omega)$.
It is found that $\sigma^{\uparrow}_\pm (\omega)\neq \sigma^{\downarrow}_\pm (\omega)$ even in the normal state indicating the possibility of the spin-selective excitation depending on the handedness of the circularly polarized light.
For $h<h_{\rm c}$, the peak structure is observed in $\sigma^\uparrow_+(\omega)=\sigma^\downarrow_-(\omega)$ at $\omega=\Delta_1$, while such a peak is not observed in $\sigma^\uparrow_-(\omega)=\sigma^\downarrow_+(\omega)$.
The excitation energy $\omega=\Delta_1$ corresponds to the smallest direct band gap at K- and/or $-$K-points.
In the CO state, since the band gap is smallest at $-$K (K)-point for spin-$\uparrow$ ($\downarrow$) electron [see Fig.~\ref{fig_band}(a)], the same peaks of $\sigma^{\uparrow}_{+}(\omega)$ and $\sigma^{\downarrow}_{-}(\omega)$ originate from the excitation at different $\bm{k}$ points, namely, the $-$K and K points, respectively [see also Fig.\ref{fig_intensity}(b)].
In contrast, in the $z$SO state, the peak structures in $\sigma^{\uparrow}_{+}(\omega)$ and $\sigma^{\downarrow}_{-}(\omega)$ are due to the excitation at the same $-$K-point.
It is also understood by the PH transformation, namely, for $\downarrow$ spin the role of K point in the CO state is converted to $-$K point in the $z$SO state.
With increasing the molecular field $h$, the peak in $\sigma^\uparrow_+(\omega)=\sigma^\downarrow_- (\omega)$ at $\omega=\Delta_1$ shifts to lower energy for $h<h_{\rm c}$.
With further increasing $h$ above $h_{\rm c}$, the peak in $\sigma^\uparrow_+(\omega)=\sigma^\downarrow_-(\omega)$ collapses, while the corresponding peak appears in $\sigma^\uparrow_-(\omega)=\sigma^\downarrow_+(\omega)$.
These results clearly display the sudden change of the optical selection rules  across the topological transition point $h=h_{\rm c}$.

In order to get clear insight into the optical selection rules, we investigate the momentum-resolved intensities of the excitation from the valence state  $|\bm{k} \sigma \nu=2 \rangle$ to the conduction state  $|\bm{k} \sigma \nu'=3 \rangle$, $P^{\sigma}_{\pm}(\bm{k})$, given by
\begin{align}
P^{\sigma}_{\pm}(\bm{k})= \frac{| \langle  \bm{k}\sigma 3| j_{\pm}^{\sigma} |\bm{k} \sigma 2 \rangle|^2}{\varepsilon_{\bm{k}\sigma 3}- \varepsilon_{\bm{k}\sigma 2}}.
\end{align}
As shown in Fig.~\ref{fig_intensity}(a), in the normal state, $P^{\uparrow}_{+}(\bm{k})$ and $P^{\downarrow}_{-}(\bm{k})$ exhibit the prominent intensities at K- and $-$K-points, and all these intensities at K and $-$K points are the same.
On the other hand, $P^{\downarrow}_{+}(\bm{k})$ and $P^{\uparrow}_{-}(\bm{k})$ are relatively small in the whole region of the Brillouin zone.

After switching on the molecular field, breaking of the inversion symmetry leads to inequivalence of $P_{\pm}^{\sigma}(\bm{k})$ between K and $-$K points as shown in Figs.~\ref{fig_intensity}(b) and (c).
This is a clear signature of the spin-valley selective excitation with respect to the absorption of the circularly polarized light with frequency $\omega=\Delta_1$.
In the CO state for $h=0.1<h_{\rm c}$, it is found that $P^{\uparrow}_{+}(-\bm{K})=P^{\downarrow}_{-}(\bm{K})$ become much larger than $P^{\uparrow}_{+}(\bm{K}) = P^{\downarrow}_{-}(-\bm{K})$.
This indicates the following spin-valley selectivity in the optical absorption: excitation of the electron with spin $\uparrow$ ($\downarrow$) at $\bm{k}=-\bm{K}$ ($\bm{K}$) by the right (left) circularly polarized light absorption.
$P^{\sigma}_{\pm}(\bm{k})$ continuously varies as a function of $h$ up to $h=h_{\rm c}$, and the discontinuous change occurs across the topological transition point $h=h_{\rm c}$.
For $h=0.4>h_{\rm c}$, $P^{\downarrow}_{+}(\bm{K})=P^{\uparrow}_{-}(-\bm{K})$ become much larger than $P^{\downarrow}_-(\bm{k})=P^{\uparrow}_+(\bm{k})$ in contrast to the case of $h<h_{\rm c}$.
In other words, the optical selection rules change abruptly across the topological transition point $h=h_{\rm c}$ (see  Table~\ref{selection-rule}).

As shown in Fig.~\ref{fig_intensity}(c), the sudden change of the optical selection rules are realized also in the $z$SO states although the spin-valley selectivity is different from that of the CO state.
For $h<h_{\rm c}$, the selective excitation of the electron with spin $\uparrow$ ($\downarrow$) and wave vector $-\bm{K}$ by the right (left) circularly polarized light is possible, while for $h>h_{\rm c}$, the electron with spin $\downarrow$ ($\uparrow$) at $-\bm{K}$ point is selectively excited by the right (left) circularly polarized light.
The above-mentioned optical selection rules are summarized in Table~\ref{selection-rule}.

\begin{figure}[t]
\begin{center}
\includegraphics[width=8cm]{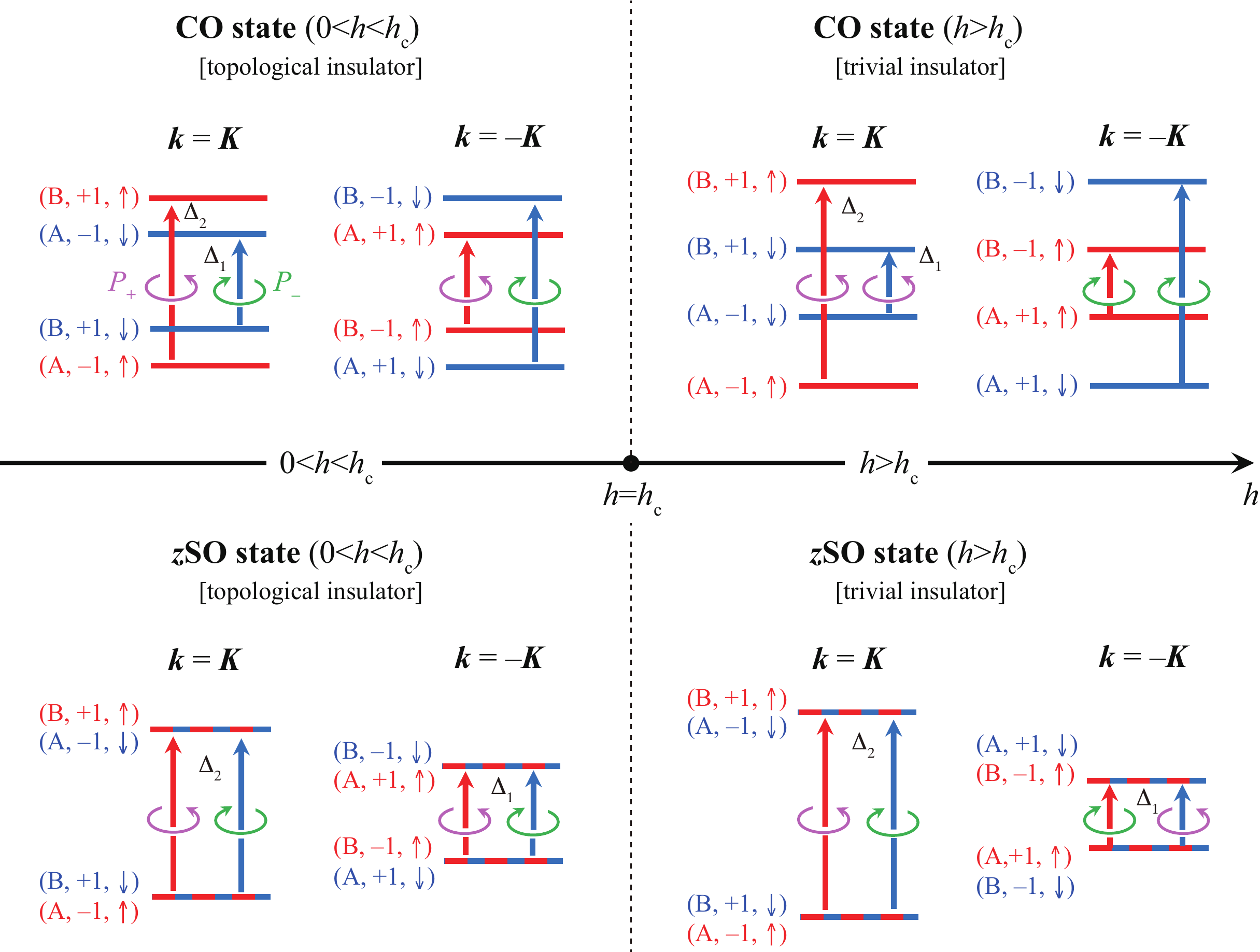} 
\caption{
(Color online) Schematic illustration of the energy levels and optical selection rules.
$P_{+(-)}$ represents the right (left) circularly light absorption at the excitation energies $\Delta_1$ and $\Delta_2$.
The relevant eigenstates for the spin-valley selective absorption at $\bm{k}=\pm\bm{K}$ are indicated by $(s,m,\sigma)$.
The relation for the $z$SO state may be obtained by replacing the energy levels of $(s,m,\downarrow)$ at $\bm{k}=\bm{K}$ in the CO state by the opposite sign of those of $(s,-m,\downarrow)$ at $\bm{k}=-\bm{K}$.
}
\label{fig_selection}
\end{center}
\end{figure}

The analysis of the low-energy part of the Hamiltonian and the current operator at $\bm{k}=\pm \bm{K}$ is helpful to understand the optical properties.
It is easily verified that
 in terms of four basis, $|\mathrm{A},\mp1,\sigma\rangle$ and $|\mathrm{B},\pm1,\sigma\rangle$ ($\sigma=\uparrow$, $\downarrow$), the low-energy part of 
the Hamiltonian at $\bm{k}=\pm\bm{K}$ is diagonalized.
The eigenenergies of $|\mathrm{A},\mp1,\sigma\rangle$ and $|\mathrm{B},\pm1,\sigma\rangle$ in the CO state are given by $-h\mp h_{\rm c}\sigma$ and $h\pm h_{\rm c}\sigma$, respectively.
The eigenenergies in the $z$SO state can be obtained by replacing $h$ with $h\sigma$.
In terms of the basis, $(|\mathrm{A},\mp 1,\sigma\rangle,  |\mathrm{B},\pm1,\sigma\rangle )$, the non-zero component of the current operator at $\bm{k}=\pm\bm{K}$ is given by
$\langle \mathrm{B},\pm1,\sigma|j^\sigma_{\pm}|\mathrm{A},\mp1,\sigma\rangle=i\sqrt{3}t_{1}$.
The structure of the current operator and the $h$-dependence of the energy levels explain the abrupt change of the optical selection rules at the topological transition point.
Figure~\ref{fig_selection} schematically shows the energy levels and the optical selection rules.
The CO state is considered in the following, while the analysis for the $z$SO state is almost parallel to the case of the CO state.
For $h<h_{\rm c}$, the spin $\downarrow$ ($\uparrow$) states form the conduction and valence states at K- ($-$K-) point in the vicinity of the Fermi energy as shown in Fig.~\ref{fig_selection} (upper-left panel), and thus, the intensities $P^{\sigma}_{\pm}(\bm{k})$ at $\bm{k}=\bm{K}$ and $-\bm{K}$ are obtained as follows,
\begin{align}
&
P^{\uparrow}_{+}(-\bm{K}) = P^{\downarrow}_{-}(\bm{K}) =\frac{3t_1^2}{\Delta_1},
\quad
P^{\uparrow}_{+}(\bm{K}) = P^{\downarrow}_{-}(-\bm{K}) =\frac{3t_1^2}{\Delta_2},
\cr
&
P^{\uparrow}_{-}(\pm\bm{K}) = P^{\downarrow}_{+}(\pm\bm{K}) =0.
\label{eq_pa}
\end{align}
With increasing $h$, the energy splitting $\Delta_1=2|h-h_{\rm c}|$ between the valence and conduction states for spin $\downarrow$ ($\uparrow$) at $\bm{k}=\bm{K}$ ($-\bm{K}$) gets smaller, and the level crossing takes place across the topological transition point $h=h_{\rm c}$ as shown in Fig.~\ref{fig_selection} [see also Fig.~\ref{fig_band}(a)].
Then, for $h>h_{\rm c}$ (upper-right panel), the intensities $P^{\sigma}_{\pm}(\bm{k})$ at K- and $-$K-points are given by
\begin{align}
&
P^{\uparrow}_{-}(-\bm{K}) = P^{\downarrow}_{+}(\bm{K}) =\frac{3t_1^2}{\Delta_1},
\quad
P^{\uparrow}_{+}(\bm{K}) = P^{\downarrow}_{-}(-\bm{K}) =\frac{3t_1^2}{\Delta_2},
\cr
&
P^{\uparrow}_{\pm}(\mp\bm{K}) = P^{\downarrow}_{\pm}(\mp\bm{K}) =0.
\label{eq_pb}
\end{align}
The difference between $P^{\sigma}_{\pm}(\bm{k})$ for $h<h_{\rm c}$ and for $h>h_{\rm c}$ shown in Eqs.~(\ref{eq_pa}) and (\ref{eq_pb}) accounts for the discontinuous change of the optical selection rules at $h=h_{\rm c}$.
Similarly, according to the PH transformation of Eq.~(\ref{shibatrans}), the analysis for the $z$SO state (lower panel of Fig.~\ref{fig_selection}) may be obtained by replacing the eigenenergies of $|s,m,\downarrow\rangle$ at $\bm{k}=\bm{K}$ by the opposite sign of those of $|s,-m,\downarrow\rangle$ at $\bm{k}=-\bm{K}$.

\begin{table}[t]
\caption{
Optical selection rules with respect to the light absorption with frequency $\omega=\Delta_1$ in the CO and $z$SO states.
$(\bm{k},\sigma)$ indicates that the electron with spin $\sigma$ at $\bm{k}$ point is selectively excited by the circularly polarized light.
}
\label{selection-rule}
\vspace{2mm}
\begin{tabular}{ccc|ccc} \hline\hline
\multicolumn{3}{c|}{$0<h<h_{\rm c}$} & \multicolumn{2}{c}{$h>h_{\rm c}$} \\ \hline
       & right & left & right & left \\ \hline
CO     & $(-\bm{K},\uparrow)$ & $(\bm{K},\downarrow)$ & $(\bm{K},\downarrow)$ & $(-\bm{K},\uparrow)$ \\
$z$SO  & $(-\bm{K},\uparrow)$ & $(-\bm{K},\downarrow)$ & $(-\bm{K},\downarrow)$ & $(-\bm{K},\uparrow)$ \\ \hline\hline
\end{tabular}
\end{table}

In summary, we have revealed the optical properties of the two-orbital model in the ordered state accompanying spontaneous inversion symmetry breaking.
Our results demonstrate that the spin-valley selective excitation by the circularly polarized light is possible in the charge and $z$-type antiferromagnetic ordered states, and the distinct optical selection rules are found in each ordered phases.
The CO and $z$SO states in our model provide similar situations dynamically as the monolayer MoS$_2$ in the paramagnetic state~\cite{Xiao_2} and MnPX$_3$ (X=S, Se) in the antiferromagnetic state~\cite{Li}, where the spin-valley coupled optical properties have been previously studied.
In the present study, the intuitive understanding of the relation between those two situations is obtained by applying the particle-hole transformation to a minimal two-orbital model.
It is also found that optical selection rules depend on the topological properties of insulating band structures, which was
first pointed out in the context of silicene under the external electric field~\cite{Ezawa_1}.
Our results indicate that essentially the same physics could be realized as the spontaneous symmetry breaking state, and it may enhance further the spin-valleytronics applications in a wide class of centrosymmetric materials with local asymmetry sites.

\begin{acknowledgments}
The authors would like to thank S. Hayami, Y. Motome, and T. Takimoto for fruitful discussions.
This work has been supported by JSPS KAKENHI Grant Number 15H05885 (J-Physics) and 15K05176.
\end{acknowledgments}

\end{document}